\documentstyle[aps,prl,preprint]{revtex}
\begin{document}
\title {Shadow bands in overdoped $Bi_2Sr_2CaCu_2O_{8+x}$}
\author{S. La Rosa$^{1}$, R.J. Kelley$^{2}$, C. Kendziora$^{3}$,
 G. Margaritondo$^{1}$, M.
Onellion$^{2}$ and A. Chubukov$^{2,4}$}
\address{$^{1}$ Institut de Physique Appliquee,
Ecole Polytechnique Federale Lausanne, Switzerland\\
$^{2}$ Physics Department, University of Wisconsin, Madison, WI 53706\\
$^{3}$ Naval Research Laboratory, Washington, D.C.\\
$^{4}$ P.L. Kapitza Institute for Physical Problems, Moscow, Russia}
\date{today}
\maketitle

\begin{abstract}
We performed angle-resolved photoemission experiments on
overdoped single crystal $Bi_2Sr_2CaCu_2O_{8+x}$ samples with
Tc values as low as 55K and very narrow transitions as measured by
AC susceptibility.
The photoemission data indicate the presence of
a ``shadow'' peak in the quasiparticle spectral function $A(\omega)$
shifted by $(\pi,\pi)$
from the conventional quasiparticle peak. The ratio of the
shadow/quasiparticle peak amplitudes strongly
increases with the binding energy. This is consistent with the idea
that the shadow peak is not a structural replica of the
quasiparticle peak, but rather a
maximum in a spectral function which can exist due to interaction with spin
fluctuations peaked at $(\pi,\pi)$.
\end{abstract}


The question of what boson mediates the superconducting pairing interaction is
of great interest in high-temperature superconductors.
Alternatives to phonons include spin fluctuations~\cite{para}.
The exchange by spin fluctuations peaked at antiferromagnetic momentum $Q =
(\pi,\pi)$ yields  a $d-$wave pairing state which is
consistent with a number of experiments on high-$T_c$ materials. However,
there are restrictions which have to be met in order to have strong
attraction in a $d-wave$ channel: (i)
 spin fluctuations should be  peaked at or very near  $Q$; (ii)
one should be able to find a region of fermionic momentum
along the directions where the $d$-wave gap has a maximum, where
$k$ and $k+Q$ are both located near the Fermi surface; (iii) in the
density of states, there should be
no precursors of the upper and lower bands of the magnetically ordered state -
otherwise vertex corrections would strongly decrease pairing
interaction~\cite{schr}. Near
optimal doping, neutron scattering~\cite{neutrons},
and photoemission experiments~\cite{photoemiss} and numerical studies  of the
density of states~\cite{dagotto}
 indicate that all three conditions are satisfied.

An open issue
is  whether the weak peak at $k =Q$
in the dynamical susceptibility observed at
optimal doping means that spin fluctuations are strong
enough to
yield $T_c$ values consistent with experiment.   In photoemission studies,
a way to check whether or not spin fluctuations are
relevant is to study the form of $A(k,\omega)$ for $k \approx k_F +Q$
where $k_F$ is the (angle dependent) Fermi momentum.
If spin fluctuations at optimal doping are still strong, then
$A(k)$ measured at a given {\it small} frequency should have a second maximum
at  $k = k_F +Q$ due
to singular self-energy corrections~\cite{Chu95}.
An important distinction is that
this second peak in $A(k)$ is not a second quasiparticle peak but just an
enhancement of the low-frequency part of the spectral function. In this sense,
the situation at optimal doping is qualitatively different from that at and
very near to half-filling where the valence and
conduction bands are almost preformed. In the latter case,
$A(\omega)$ measured at $k \approx k_F$ and $k_F + Q$
should behave nearly in the same
way, i.e., the photoemission data should reveal two {\it
quasiparticle} peaks separated by $Q$~\cite{ks}.
In fully doped materials, however,
the spectral weight is transformed from the conduction to the valence band,
and the second quasiparticle peak should
disappear; when and how this
happens, however, depends on the details of the electronic structure.

The momentum dependence of the spectral function in $Bi2212$
at the smallest possible
frequency has been measured by Aebi et al.~\cite{aebi}. They reported
observing a second, shadow peak, but have been careful
to note
 that it is an open question whether this peak  to due to
antiferromagnetic fluctuations or it merely
reflects the presence of the structural replica of the Fermi surface and
the quasiparticle band.

In this communication we
report on measurements of  the relative
intensity of the shadow peak and the main quasiparticle
peak  as a function of the binding energy. These measurements
distinguish between magnetic and structural
origins of the shadow
peak. The reason is as follows.  If the
shadow peak is the structural replica of the quasiparticle peak,
 it should have the same Fermi-liquid
frequency dependence at the lowest frequencies, $A(\omega) \propto
\omega^{-2}$, as the main peak, and the ratio of the two
should be virtually independent on frequency. If, on the contrary, the second
peak is
just an enhancement of the low frequency part of $A(\omega)$ at $k_F +Q$ due to
spin fluctuations, then the relative intensity of the shadow peak should
vanish when binding energy tends to zero.

We have studied overdoped $Bi_2Sr_2CaCu_2O_{8+x}$
single crystals, with transition temperatures, Tc=55-62K and
report observing shadow peaks  using conventional angle-resolved
photoemission spectroscopy in which one measures spectral function as a
function of frequency at a given momentum.
  We found that the relative intensity of the
shadow peak strongly increases with increasing binding energy which agrees
 with the idea that the shadow peak is not a second quasiparticle
peak.

Single crystal samples of $Bi_2Sr_2CaCu_2O_{8+x}$ were grown using a self-flux
technique in a strong horizontal thermal gradient to stabilize the direction of
solidification.[10,11] The individual crystals were manually separated from the
bulk charge utilizing their propensity to cleave micaciously perpendicular to
the crystallographic $c$-axis.
Typical dimensions for samples in this study were
$4mm \times 4mm \times 30um$, with the shortest length along the $c$-axis.

Our AC susceptibility measurements have shown that
the crystals exhibit a $10-90\%$ transition in a $T$ range of $0.5K$ for both
optimally doped and overdoped samples.
Previous reports~\cite{16} indicated
 a best transition width of $1.3K$ for optimally
doped material and $~3K$ for somewhat overdoped samples (Tc=72-75K). The
narrow transition width indicates a more homogeneous oxygen and cation doping
than in earlier studies.
 Fig.~\ref{fig_one}
 illustrates AC susceptibility measurements of overdoped and
optimally doped $Bi_2Sr_2CaCu_2O_{8+x}$
single crystals. There is no indication of either $Bi_2Sr_2CuO_{6+x}$
or $Bi_2Sr_2CaCu_3O_{10+x}$
phases. This is confirmed by x-ray diffraction and resistivity measurements, as
reported in detail elsewhere~\cite{diffr}.

Angle-resolved photoemission measurements were performed using a $50 mm$ radius
hemispherical electron energy analyzer, mounted on a two-axis goniometer. The
samples were transferred from a load lock chamber to the ultrahigh vacuum
photoemission chamber (base pressure 6-7E-11 torr). The samples were cooled to
just above their superconducting transition temperature of $55-62K$ and cleaved
while cold. The sample orientation with respect to the sample surface normal
was adjusted in situ and confirmed using low energy electron
diffraction~\cite{diffract}

Fig. \ref{fig_two}
illustrates angle-resolved photoemission spectra taken along the
$\Gamma-X (k_x=k_y)$ and $\Gamma -Y (k_x = -k_y)$ directions in
the Brillouin zone, for both cases we plotted the data
for an overdoped sample with $T_c=55K$.
 Several noteworthy points emerge from the data. First,
the main Fermi surface
crossing, at $(0.38 \pm 0.01) Q$, occurs at a slightly smaller wavevector than
that
for optimally doped or slightly overdoped
samples~\cite{17}, because our samples have a higher carrier concentration.
Second, we found another
peak in the photoemission intensity,  that appears at our smallest frequency
at $(0.64 \pm 0.02)Q$. Within our experimental error, this second peak is
located exactly at a distance $Q =(\pi,\pi)$ from the Fermi surface, i.e., it
is a shadow peak.
As the wavevector is increased further from $0.64 Q$,
the shadow peak  disperses towards higher binding
energies. By a wavector of $0.75 Q$, the peak, now very weak, is at a
binding energy of 180-200 meV.

We reproducibly observed these new features of the photoemission spectra
on three different samples and
in both the $\Gamma-X$ and $\Gamma-Y$ directions.
Including looking at more than one
angular cut on some samples, we looked for, and found shadow peaks
in a total of eight angular cuts.
Our results obtained on different samples
are presented  in Fig. ~\ref{fig_three}.  The main Fermi surface
measured by P. Aebi et.al. is shown by a solid line,
 the location of the shadow
peak by a dotted line, and our
experimental results by open circles.
 The Fermi surface wavevectors obtained
for all three samples are, within our experimental error,
the same, so the same
open circle symbol is used for all samples.
 The strength of the intensity at the shadow peaks is
polarization-dependent; the details of this will be reported
elsewhere~\cite{18}.

Fig. \ref{fig_four}
illustrates the ratio, $R$, of the shadow
peak/ main quasiparticle peak amplitudes as a function
of binding energy.  Data were obtained from six angular cuts on four samples.
The ratio increases as the binding energy increases;
the increase is as much as by a factor of $4.2$.
It is very unlikely that such increase  can be explained if one assumes that
the shadow peak is just a structural replica of the quasiparticle peak.
A more realistic conclusion from the data is
that there is just one Fermi surface, located at $k_F$, one band of
propagating quasiparticle excitations near $k_F$, while the shadow peak is
just an anomaly in the spectral function. This conclusion also agrees with
our experimental observation that
 the increase of the shadow peak/main peak amplitude
ratio is primarily due
to the decrease of the main peak amplitude as binding energy increases;
the amplitude of the shadow peak demonstrates only weak dependence
on the binding energy.

As we already mentioned,
the existence of the second peak in $A(k, \omega \rightarrow 0)$
 located at $k_F +Q$, can be
explained as due to the interaction with spin fluctuations.
 The exchange of
spin fluctuation yields fermionic self-energy $\Sigma (k,\omega) =g^2 \int
d{\vec q} d\Omega G_0 (k+q, \omega + \Omega)~\chi (q,\Omega)$, where
 $g$ is the coupling constant,
 $G_0 (k,\omega) = \omega - (\epsilon_k - \mu) + i \delta
sgn \omega$ is the Green function for free fermions ($\epsilon_k$ is an
excitation energy and $\mu$ is the chemical potential).
The magnetic susceptibility $\chi (q,\Omega)$
 describes overdamped spin excitations and is assumed to be peaked
at $Q$. The low-energy part of $\chi (q,\Omega)$ is well approximated by
$\chi^{-1} (q,\Omega) \propto (\Delta^2 + v^2 (q-Q)^2 + 2 i \gamma
\Omega)$~\cite{david}, where $\gamma$ (the damping term) and $v$ are of the
order of the electronic bandwidth, $W$, and  $\Delta$ is the gap which
vanishes at the antiferromagnetic transition point. The near antiferromagnetism
implies that $\Delta \ll W$.

Consider now a region of $k$ values which are
 close to the shadow point $k_F +Q$, but are relatively far from $k_F$ such
that $\epsilon_k - \mu = O(W)$.
In this situation, $A (k, \omega) \propto Im \Sigma (k, \omega)$. Now, the key
point is that the self-energy is nearly singular
for $k = k_F +Q$ and $\Delta \ll W$, and this near
singularity gives rise to a maximum in $A(k)$ at $k = k_F +Q$.
The full expression for the
low-frequency part of $\Sigma (k, \omega)$ was obtained in ~\cite{Chu95,mil},
 and here we
present only the results for several limiting cases.
For $\Delta =0$, we have
$$A(k, \omega) \propto \frac{\omega^{2}}{(\epsilon_{k+Q} -\mu)^3}$$
 for
$\omega < (\epsilon_{k+Q} -\mu)^2/\omega_0$ ($\omega_0 = O(W)$),
 and
$$A(k, \omega) \propto
\omega^{1/2}$$
 for $\omega > (\epsilon_{k+Q} -\mu)^2 /\omega_0$.
Clearly, $A(k)$, measured at a given, small $\omega$ has a maximum when
$\epsilon_{k+Q} = \mu$, i.e., $k = k_F +Q$. For $\Delta \neq 0$, the behavior
is more complex, but the peak in $A(k)$ at $k = k_F + Q$ survives as long as
$\Delta \ll W$.

The above arguments can be applied to the experiments of Aebi et al., who
measured $A(k)$ at a given frequency.
 The present experiments were performed in a different way: the
photoemission
spectra were obtained as functions of the binding energy (i.e., frequency) for
several values of $k$. The location of the shadow peak was obtained as the
position of the maximum of $A(\omega)$ at a given $k$.
To account for these data, the low-energy spin-fluctuation
approach has to be modified because in the present form,
it allows one to predict only the form of $A(\omega)$ when
it increases as a function of frequency.
At larger frequencies, $A(\omega)$
 indeed passes through a maximum and then  decreases.
However, to find the self-energy in this frequency range,
one has to utilize the sum rule for the susceptibility which in turn
requires one either
to add higher powers of frequency and momentum into the expression for
susceptibility or introduce
a cutoff in the momentum/frequency integration over bosonic variables. In both
cases, the maximum of $A(\omega)$ is expected to be located at
some frequency compared to
$\omega_0$ which apparently
does not depend crucially on how close $k$ is to $k_F+Q$.
The amplitude of $A(\omega)$ near the maximum  is also
expected not to depend strongly on $\epsilon_{k+Q} - \mu$.  In other words,
within weak coupling spin-fluctuation approach (when coupling is small enough
to yield a pseudogap),  only the
low-frequency part of $A(\omega)$ changes drastically around $k= k_F +Q$,
while the  rest of $A(\omega)$ doesn't change much.
The near independence of the shadow peak amplitude on the binding energy is
consistent with experiment - the shadow peak amplitude does
demonstrate some frequency dependence with the maximum at $\omega =
70-90 meV$, but the maximum observed change in the amplitude is much less than
that of the main band, by a factor of
(x3.3) or greater.
In Fig.~\ref{fig_four} we fitted the data for the amplitude ratio by an
$\omega^2$
dependence which would be the case if the
shadow peak was completely $\omega$
independent, while the main peak follow the
Fermi-liquid prediction $A(\omega =
\epsilon_k - \mu) \propto \omega^{-2}$. We see that
at small frequencies, the fit works reasonably well, but there are deviations
at higher frequencies as expected
due to higher-order terms in the
frequency expansion.

At the same time, the observed strong variation of the shadow peak location
with the deviation from $k_F - Q$ cannot presently be explained
 within the weak-coupling
spin-fluctuation approach.
One possibility is that already in
a weak-coupling spin-fluctuation theory,
the position of the peak in $A(\omega)$ in fact does have some moderate
$k$ dependence as a result of the interplay between the parameters in spin
susceptibility and electronic dispersion. Another possibility is that some
features of the strong-coupling, almost spin-density-wave solution~\cite{sdw},
specifically the existence of a second,
propagating quasiparticle pseudoband, can still be observed at not very small
frequencies even in the overdoped case.

To conclude, in this paper we reported the observation of the shadow peaks
in photoemission studies of overdoped $2212 Bi$ samples.
We studied  the
relative intensity of the shadow and main peaks and found that
it strongly increases with frequency.
This indicates that  the shadow
peak is not a structural replica of the main peak, but rather an enhancement of
the spectral function at $k \approx k_F +Q$.
This enhancement may be due to the interaction with spin fluctuations,
and our data thus provide some support to the idea~\cite{para} that
spin-fluctuations remain peaked at $(\pi,\pi)$ even in the overdoped regime.
 That is, even highly overdoped samples appear to support a paramagnon as a
possible exchange boson
involved in superconductivity.
On the other hand, the shadow peak exhibits a strong
$k$-dependence of the peak position. Within the spin-fluctuation model, we
cannot presently provide a definitive explanation for this observation.

We benefitted from conversations with Philippe Aebi, Ivan Bozovic, Robert
Joynt and Igor Mazin. Financial support was provided by the U.S. NSF,
both directly and
through support of the Wisconsin Synchrotron Radiation Center, Ecole
Polytechnique Federale Lausanne, and the Fonds National Suisse de la Recherche
Scientifique.  SLR is an EC Fellow,
C.K. is a National Research Council Fellow, and A.C. is an A.P. Sloan Fellow.

\begin{figure}
\caption{AC susceptibility measurements of optimally doped, as grown, and
overdoped samples. The $10-90\%$ transition
width is $0.5K$ for all samples studied}
\label{fig_one}
\end{figure}

\begin{figure}
\caption {(a)Series of photoemission energy distribution curves (EDC's) versus
location in the Brillouin zone along the $(\pi,\pi)$
 direction. Inset illustrates
location in Brillouin zone at
which each spectrum was taken. (b) Corresponding EDC's for the  $(\pi,-\pi)$
direction. Note the appearance of a shadow peak for both directions. }
\label{fig_two}
\end{figure}

\begin{figure}
\caption {Location of main Fermi surface crossing and a shadow
peak nearest to the Fermi
energy (open circles), obtained from a series of data like as in Fig. 2.
Solid line is Fermi surface for optimally doped samples.
Dotted line is shadow peak observed in Ref.\protect\cite{aebi}
 for optimally doped samples. Samples are more overdoped than
Ref.\protect\cite{aebi}, so the main crossing is closer to the center of the
Brillouin zone. The shadow peaks are shifted from the main peaks by
$(\pi,\pi)$.}
\label{fig_three}
\end{figure}

\begin{figure}
\caption {Ratio $R$ of shadow peak/main quasiparticle peak amplitudes
 as a function of binding energy $E_b$, taken from a series of
data like as in Fig. 2.
The ratio $R$ has been normalized to $1.0$
 at lowest binding energy (45 meV). Open circles
are data taken on three different samples. Dotted line is the result expected
if the shadow peak is structural in origin. Solid line is the
result expected if
the shadow peak is due to short-ranged spin fluctuations.}
\label{fig_four}
\end{figure}

\end{document}